\documentclass[preprint,showpacs,preprintnumbers,amsmath,amssymb]{revtex4}

\usepackage{graphicx}
\usepackage{dcolumn}
\usepackage{bm}
\begin{document}
\title{Transport through short quantum wires }
\author{D. Schmeltzer}

\affiliation{Physics Department \\ City College of the City
University of New York
\\ New York, New York 10031}

\date{\today}
\begin{abstract}
At temperatures $T < {\hbar v_F}/{K_B d} \equiv T^{wire}$ the
collective excitations are negligible and the spectrum of the
short wire is dominated by the ''zero modes'' particle
excitations.  At temperature $ T > T^{wire}$ a spin polarized
state controlled by the electron-electron interaction and electron
density is identified at short times. As a result the anomaly in
the conductance $G > 2 \times 0.5 {e^2}/{h} $ appears. At $T
\rightarrow 0$ by varying the gate voltage we find that our
problem is equivalent to a resonant impurity level. As a result
perfect transmission with a conductance $G \simeq \frac{2 e^2}{h}$
is obtained. The model presented here can be used as a spin filter
which operate by varying the temperature. Transport through a
short quantum wire with electron-electron interaction and length
``$d$'' coupled to Luttinger leads is considered. The short wire
model might explain the ``0.7 anomaly'' observed in quantum point
contacts.
\end{abstract}

\maketitle
Recent experiments in Quantum Wires show that the spin degrees of
freedom play a crucial role in the ''Quantum Ballistic
transport''. Few years ago the Cavendich group [1] reported a new
quasiplateau with the conductance $G \approx 2 \times 0.7
\frac{e^2}{h}$.  With increase in the external magnetic field, the
new quaiplateau shifts to lower values approaching $G = 2 \times
0.5 \frac{e^2}{h}$ which corresponds to the conductance of a
single polarized channel.  An attempt to explain these results
have been presented at the Ban-Ilan Conference (1997) [2].  The
explanation was based on the fact that for short quantum wires a
spin polarized state can occur causing the appearance of the new
quasiplateau. Additional explanation based on spin-polarization
have been given in references [3-5]. On the experimental side the
additional quaiplateau have been observed by a number of groups
[6-8].  The additional quasiplateau has been also observed in a
Quantum point contacts (Q.P.C.) experiment [9] suggestive of the
Kondo effect [10]. \\

Recently a density spin polarization in ultra-low disorder quantum
wires has been observed [11].  In particular it has been reported
that the feature in the range $0.5 - 0.7 \times \frac{2 e^2}{h}$
conductance depends on the electron density, length of the quantum
wire and temperature. In order to explain these results we will
introduce the following model.  We consider a short wire with
electron-electron (e-e) interaction and length $2 d$ described by
the hamiltonian $H^{wire}$.  The short wire is coupled to the left
and right leads of length $L$, $H^{leads} = H^L + H^R$ ($H^L$ and
$H^R$ represent the left and right Luttinger liquid).  The
coupling between the leads and the wire is described by the
tunneling hamiltonian $H_T$.  The conductance is determined by the
properties of $H^{wire}$.  We find that when the wire is short the
e-e interaction causes the appearance of a spin polarized state
for temperatures $ T^{wire} < T $. At $ T < T^{wire} $ we obtain a
problem which is equivalent to a resonant impurity problem.
Varying the gate voltage we reach perfect transmission.  Therefore
at $T \rightarrow 0$ one obtains a perfect conductance $G = 2
e^2/h$. At $T > T^{wire}$ we find that at short time the tunneling
electrons see the short wire polarized causing a conductance $G >
2 \times 0.5 \frac{e^2}{h}$. \\

These results are obtained within the zero mode bosonization [12].
In particular, we find that for short wire the backward scattering
amplitude for the sine-Gordon spin liquid does not normalize to
zero as is the case for long wire.  As a result a spin
polarization occurs.  This new feature is obtained within a new
non-linear zero mode calculation.  The conductance is determined
by the zero mode correlation function of the short wire $ K_\sigma
(t - t_1) = \langle e^{i q_\sigma (t)} e^{- i
q_\sigma(t_1)}\rangle $ where $q_\sigma$, $\sigma = \uparrow,
\downarrow$ is the zero mode coordinate $ - \pi \leq q_\sigma \leq
\pi$ conjugated to the number of fermion $N_\sigma$, $\sigma =
\uparrow, \downarrow$ in the wire.  At $T \rightarrow 0$, the two
correlation functions $K_\uparrow(t-t_1) = K_\uparrow(t-t_1)$ are
equal giving rise to a conductance $G\approx 2 e^2/h$, contrary to
finite temperatures where the spin polarized wire gives
$K_\uparrow(t - t_1) \neq K_\downarrow(t-t_1)$ causing the new
feature in the conductance $G
> 2 \times 0.5 e^2 /h$.  The range of the temperatures where this
is obtained is determined by
the length of the wire and electron density. \\

In the remaining part we are going to present our model and
results,
\begin{equation}
H = H^L + H^R + H^{wire} + H_T
\end{equation}
$H^L$ and $H^R$ represent the left and right Luttinger leads
characterized by the tunneling exponent [12, 13], $r =
\frac{1}{2}(\frac{1}{K_c} + \frac{1}{K_s}) \geq 1$ where $K_c$ and
$K_s$ are the charge and spin interaction parameters. \\

The leads are confined to the region $ -L \leq x \leq - d$ and $d
\leq x \leq L$, where $L \gg d > a$, ''$a$'' being the lattice
constant and ''$2d$'' the length of the short wire described by
the hamiltonian $H^{wire}$.  The ''wire'' is coupled to the leads
through the tunneling matrix elements $t_L$ at $x = -d$ and $t_R$
at $x = d$.\\

We will use open boundary condition for the two leads and wire
[14, 15].  As a result the fermions in the leads will be given by
one chiral fermion, $L_\sigma(x)$ (left leads) $R_\sigma(x)$
(right leads) and $\chi_\sigma(x)$ (for the wire).  The tunneling
hamiltonian $H_T$ takes the form,
\begin{equation}
H_T = \sum_{\sigma = \uparrow, \downarrow} \left\{ \lambda_L
\left[ L_\sigma^\dag(-d) \chi_\sigma(-d) + h.c. \right] +
\lambda_R \left[ R_\sigma^\dag(d) \chi_\sigma(d) + h.c. \right]
\right\}
\end{equation}
where $\lambda_L = 4 t_L \sin(k_F^{(L)} a) \sin(k_F^{(d)} a)$,
$\lambda_R = 4 t_R \sin(k_F^{(R)} a) \sin(k_F^{(d)} a)$, $t_L \sim
t_R$, $ k_F^{(R)} \sim k_F^{(L)} \sim k_F^{(d)}$ represent the
Fermi momentum in the two leads and in the short wire. \\

The current will depend on the properties of the short wire of
length ''2d'' with e-e interaction.  Contrary to the long wires,
for short wires we are not allowed to neglect the ''Backward''
scattering amplitude for spin excitations and ''Umklapp''
scattering for the charge part, away from half filling. \\

Within the Bosonization method the ''Backward'' and ''Umklapp''
term will be described by two sine-Gordon models [16] $H^{wire} =
H^{wire}_c + H^{wire}_s$, where $H^{wire}_c$ and $H^{wire}_s$ are
the charge and spin hamiltonian.  Since we are working with a
short wire, we will use the zero modes [12].
\begin{subequations} \label{eq3:whole}
\begin{equation}
H^{wire}_c = \int_{-d}^d dx [ v_c (\partial_x \tilde{\Phi}_c(x)
)^2 + \frac{U}{(\pi a)^2} \cos( (4 k_F - G) x + \sqrt{8 \pi}
(\Phi_c(x) - \Phi_c(-x) ) + \frac{2 \pi}{d} P_c x )] + \frac{\hbar
\pi}{4 d}v_c P_c^2  \label{subeq3:1}
\end{equation}
where $P_c = N_\uparrow + N_\downarrow$, $q_c = \frac{1}{2} (
q_\uparrow + q_\downarrow)$ and $[P_c, q_c] = i$ (''$P_c$'' is the
charge and $q_c$ is the zero mode coordinate). $4 k_F - G =
\delta$ with $G= \frac{2 \pi}{a}$, $\delta = 0$ corresponds to
half filling (when $d \rightarrow \infty$ and $\delta \neq 0$ the
sine-Gordon term can be ignored.)
\begin{equation}
H^{wire}_s = \int_{-d}^d dx [ v_s (\partial_x \tilde{\Phi}_s(x)
)^2 + \frac{U}{ 2 (\pi a)^2} \cos( \sqrt{8 \pi} (\Phi_s(x)
-\Phi_s(-x) ) + \frac{ \pi}{d} (P_s + \hat{\mu}_s) x) ] +
\frac{\hbar \pi v_s}{4 d} (P_s + \hat{\mu}_s)^2 \label{subeq3:2}
\end{equation}
\end{subequations}
In equation 3b, $N_\uparrow - N_\downarrow$ and $q_s = \frac{1}{2}
(q_\uparrow - q_\downarrow)$ are the spin zero mode variables.  If
a spin polarized solution exist one has to find the difference
between spin up and spin down is nonzero $ \langle N_\uparrow -
N_\downarrow \rangle \equiv \hat{\mu}_s \neq 0$.  Therefore the
conjugate variables to $q_s$ will be $P_s$ defined by $P_s = \, :
N_\uparrow - N_\downarrow : \, \equiv N_\uparrow - N_\downarrow -
\langle N_\uparrow - N_\downarrow \rangle $.
$\tilde{\Phi}_{c(s)}(x)$ represent the renormalized bosonic
fields, $\Phi_{c(s)}(x) = \frac{K_{c(s)}^{1/2}}{2} (
\tilde{\Phi}_{c(s)}(x) - \tilde{\Phi}_{c(s)}(-x)) +
\frac{K_{c(s)}^{- 1/2}}{2} ( \tilde{\Phi}_{c(s)}(x) +
\tilde{\Phi}_{c(s)}(-x)) $ where $\Phi_{c(s)}(x)$ are the bare
bosonic fields, $\Phi_{c(s)}(x) = (\Phi_\uparrow(x) \pm
\Phi_\downarrow(x) )/ \sqrt{2}$. $K_c < 1$, $K_s \geq 1$ are the
charge and spin interaction parameters.  We will use the
Renormalization Group (R.G.) in order to investigate the long wave
behavior of eqs 3a - 3b.  The hamiltonian in eqs 3a-3b has two
different behaviors a ''high temperature'' crossover behavior (a)
for temperature such that the thermal length $L_T = \frac{\hbar
v_F}{k_B T}$ is shorter than the length of the wire $2d$, $T >
T^{wire} = \frac{\hbar v_F}{k_B d}$ and (b) the ''low''
temperature when $T^{wire} < T$. In order to investigate this
situation we will use a two cut-off renormalization group for the
hamiltonian in eqs 3a-3b. We introduce a bandwidth cutoff $v_F
\Lambda \equiv K_B T_F$. In order to compute the tunneling current
a drain-source voltage $\mu_L - \mu_R = e V_{DS}$ and the gate
source voltage $e V_G = (\mu_R + \mu_L)/2 \equiv \bar{\mu}$ will
be applied. ($\mu_L$ and $\mu_R$ are the chemical potential for
the left and right leads). \\

Using the Renormalization group (R.G.) we scale down the problem
from $\Lambda$ to $\Lambda/b_0 \equiv \Lambda_0$ where $b_0 = d /
a$.  At the scale $b_0$ our problem is equivalent to an effective
single impurity problem governed by the wire Hamiltonian.  The
integration of the Fermion in the leads gives rise to a line width
of the ``impurity level'' (single particle state) given by:
$\Gamma \sim 2 \hat{\lambda}^2$. \\

Next we bosonize the leads fermions $R_\sigma(d)$ and
$L_\sigma(-d)$; $R_\sigma(x) = \frac{1}{\sqrt{2 \pi a}} e^{i
\hat{\varphi}_\sigma(x)}$ with $\hat{\varphi}_\sigma = \sqrt{4
\pi} \varphi_\sigma + \beta_\sigma$ and $L_\sigma(x) =
\frac{1}{\sqrt{2 \pi a}} e^{i \hat{\theta}_\sigma(x)}$ with
$\hat{\theta}_\sigma(x) = \sqrt{4 \pi}\theta_\sigma +
\alpha_\sigma$.  $\alpha_\sigma$ and $\beta_\sigma$ are the zero
mode variables of the leads with the conjugate number of particles
$\hat{N}_\sigma$ and $n_\sigma$, $[\alpha_\sigma, \hat{N}_\sigma]
= [ \beta_\sigma, n_\sigma] = - i$.\\

As a result we find that the tunneling Hamiltonian is given by:
\begin{subequations} \label{eq4:whole}
\begin{equation}
h_T = - \frac{i \hat{\lambda}^2}{\hbar} \sum_{\sigma=\uparrow,
\downarrow} \eta_{R,\sigma} \eta_{L,\sigma} \int_0^t dt_1 \{
K_\sigma(t-t_1) [ e^{-i \hat{\varphi}_\sigma(t)} e^{i
\hat{\theta}_\sigma(t_1)} - e^{-i \hat{\theta}_\sigma(t)} e^{i
\hat{\varphi}_\sigma(t_1)} ] - h.c. \} \label{subeq4:1}
\end{equation}
$\eta_{R,\sigma}$, $\eta_{L,\sigma}$ are real Majorana fermions.
$K_\sigma(t-t_1)$ is given by the short wire expectation value:
\begin{eqnarray} \nonumber
K_\sigma(t-t_1) = e^{- \Gamma(t-t_1)} \hat{K}_\sigma(t-t_1), \quad
t > t_1 \\
\hat{K}_\sigma(t-t_1) = \langle \chi_\sigma(d,t)
\chi_\sigma^\dag(-d, t_1) \rangle_{wire}, \quad t > t_1
\label{subeq4:2}
\end{eqnarray}
Using the bosonic representation of the short wire Hamiltonian we
find that $\tilde{K}_\sigma(t - t_1)$, is given by:
\begin{equation}
\tilde{K}_\sigma(t-t_1) = \frac{1}{2 \pi a} \langle e^{i
q_\sigma(t)} e^{-i q_\sigma(t_1)} e^{i \sqrt{4 \pi}
\Phi_\sigma(d,t)} e^{- i \sqrt{4 \pi} \Phi_\sigma(-d,t_1)}
\rangle_{wire} \label{subeq4:3}
\end{equation}
\end{subequations}

In eq. 4c $q_\sigma(t)$ is the zero mode coordinate.  The
correlation function will be computed for the two cases:

(a) The high temperature case, $T > T^{wire}$:

Using the R.G. method we compute the form of the Hamiltonian at
the scale $b=b_T = T_F / T = \exp \ell_T$.  At this scale we find
that the two body potentials $\hat{U}_c(\ell_T)$ and
$\hat{U}_s(\ell_T)$ are given by, $\hat{U}_c(\ell_T) =
\frac{\hat{U}_c(0)}{1 + \hat{U}_c(0) \ell_T}$, $\hat{U}_s(\ell_T)
= \frac{\hat{U}_s(0)}{1 + \hat{U}_s(0) \ell_T}$ where
$\hat{U}_c(0) = \frac{U}{\pi v_c}$ ( at half filling $\delta =0$),
$\hat{U}_c \approx 0$ (for $\delta \neq 0$) and $\hat{U}_s(0) =
\frac{U}{\pi v_s}$.

For $\L_T < d$ the spectrum can be replaced by a continuum
variable, $P_c \rightarrow \hat{\mu}_c$, $P_s + \hat{\mu}_s
\rightarrow \hat{\mu}_s$.  For the charge part we have:
\begin{subequations} \label{eq5:whole}
\begin{equation}
H_c \approx \frac{\epsilon_c}{2} \hat{\mu}_c^2, \quad \delta \neq
0 \label{subeq5:1}
\end{equation}
for $H_s$ we have
\begin{equation}
H_s = \epsilon_s [ \frac{\hat{\mu}_s^2}{2} + \hat{g}_s(T) \cos(\pi
\frac{\L_T}{d} \hat{\mu}_s ) ]  \label{subeq5:2}
\end{equation}
\end{subequations}
where $\hat{g}_s(T) = (\frac{e^2}{\hbar
c})(\frac{c}{v_s})(\frac{d}{L_T})(\frac{d}{a})
\frac{1}{\pi^2}(\frac{\hat{U}}{1 + \frac{U}{\pi v_s}\ell_T})$.  In
obtaining this expression we have used $\frac{U}{a} \simeq
\frac{e^2}{a} \hat{U}$ with $\hat{U} \approx 1$, $\frac{e^2}{\hbar
c} \simeq \frac{1}{137}$, $\frac{c}{v_s} \simeq 10^2$, we find for
$K_\sigma(t-t_1)$,
\begin{subequations} \label{eq6:whole}
\begin{eqnarray} \nonumber
K_\sigma(t-t_1) &\approx& \frac{1}{2 \pi L_T} \langle e^{i q_c(t)}
e^{- i q_c(t_1)} \rangle_{H_c} \langle e^{i \sigma q_s(t)} e^{- i
\sigma q_s(t_1)} \rangle_{H_s} \\ &=& \frac{1}{2 \pi L_T}
K_c(t-t_1) K_s^{(\sigma)} (t-t_1) \label{subeq6:1}
\end{eqnarray}
where $K_c(t-t_1)$ in the presence of $(\mu_R + \mu_L)/2 = e V_G$
is given by $K_c(t-t_1) = \langle e^{-i
\frac{\epsilon_c}{\hbar}\hat{\mu}_c(t-t_1)} \rangle_{H_c} = e^{-i
\omega_c(t-t_1)}$, $\omega_c \equiv \frac{e V_G}{\hbar}$. Using
the eq. of motion for $q_s$, $\dot{q_s} = \frac{i}{\hbar} [ q_s,
H^{wire}]$ we find from eq. 5b that $K_s^{(\sigma)}(t-t_1) \approx
\langle \exp[- i \sigma \frac{\epsilon_s}{\hbar} \hat{\mu}_s
(t-t_1) ] \rangle_{H_s}$ represents the expectation value with
respect to $\hat{\mu}_s$ controlled by the Hamiltonian $H_s$ in
eq. 5b. We define the parameters, $\omega_s \equiv
\frac{\epsilon_s}{\hbar}$, $y = \sqrt{\tilde{\beta}\hat{\mu}_s}$,
$\tilde{\beta} = \beta \epsilon_s = \frac{\pi}{2 d} L_T$ and
introduce the Euclidean time, $\tau - \tau_1 = i(t-t_1)$.  We
compute the spin correlation function
$K_s^{(\sigma)}(\tau-\tau')$:
\begin{equation}
K_s^{(\sigma)}(\tau-\tau_1) = Z^{-1} \int_{-\infty}^{\infty} dy
\exp - [ \frac{1}{2} y^2 + \kappa_s(T) \cos( 2 \sqrt{\pi}
(\frac{L_T}{d})^{1/2} y) + \sigma \frac{\omega_s}{\sqrt{\pi}}
(\frac{2 d}{L_T})^{1/2} y (\tau - \tau_1) ] \label{subeq6:2}
\end{equation}
where $Z = \int dy e^{- \beta H_s}$ and $\kappa_s(T) =
\hat{g}_s(T) \pi L_T / 2d$.  In order to compute the integral in
eq. 6b we use the saddle point method.  The saddle point solution
of eq. 6b is given by:
\begin{equation}
y \equiv \bar{y} = - \sigma \frac{\omega_s}{\sqrt{\pi}} (\frac{2
d}{L_T})^{1/2}(\tau - \tau_1) + \bar{x}_0 \label{subeq6:3}
\end{equation}
A polarized solution exists if $\bar{x}_0 \neq 0$.  $\bar{x}_0$ is
the solution of the eq. 6d
\begin{equation}
\bar{x}_0 - \kappa_s(T) 2 \sqrt{\pi} (\frac{L_T}{d})^{1/2} \sin[2
\sqrt{\pi} (\frac{L_T}{d})^{1/2} \bar{x}_0 - \sigma 2
\omega_s(\tau-\tau_1)] = 0 \label{subeq6:4}
\end{equation}
From eq. 6d we see that a non zero solution for $x_0$ exists if $2
\omega_s(\tau-\tau_1) \ll 1$ and is given by:
$sinc(2\sqrt{\pi}(L_T/d)^{1/2} \bar{x}_0 ) \equiv \hat{g}_s(T)
\pi(L_T/2d)$.  Such a solution is possible since the conductance
is determined by the cutoff $\tau-\tau_1 \leq \frac{L_T}{v_s}$.
This gives $\omega_s(\tau-\tau_1) \leq \frac{L_T}{d} <1$.
Therefore, in spite of the fact that no real breaking of symmetry
occurs, for $\tau-\tau_1 \leq \frac{L_T}{v_s}$ we have
$K_s^{(\sigma)}(t-t_1) \simeq e^{-
\frac{\omega_s^2}{\pi}(\frac{2d}{L_T})(t-t_1)^2} e^{-i \sigma
(\frac{\omega_s \bar{x}_0}{\sqrt{\pi}})(\frac{2d}{L_T})^{1/2}
(t-t_1)}$. As a result we find that $K_\sigma(t-t_1)$ in eq. 6a is
equivalent to impurity in a magnetic field:
\begin{eqnarray}
K_\sigma(t-t_1) &\approx& e^{-i \frac{E_\sigma}{\hbar}(t-t_1)};
\quad (t-t_1) < \frac{L_T}{v_s} \label{subeq6:5} \\
E_\sigma &=& E_c + \sigma \Delta_s; \quad \sigma = \pm
\label{subeq6:6}
\end{eqnarray}
where $E_c = \hbar \omega_c$, $\Delta_s \equiv
\frac{\omega_s}{\sqrt{\pi}}(\frac{2d}{L_T})^{1/2} \bar{x}_0$.
Using eqs. 6e and 4a we compute the tunneling current
\end{subequations}

The tunneling current is given by the difference of charge between
the left and right leads, $ I_\sigma = e \frac{d n_\sigma}{d t} =
- e \frac{\hat{N}_\sigma}{d t} = \frac{1}{2} e \frac{d J_\sigma}{d
t} $, $J_\sigma = n_\sigma - \hat{N}_\sigma = i(\frac{d}{d
\alpha_\sigma} - \frac{d}{d \beta_\sigma} ) \equiv 2 i \frac{d}{d
\gamma_\sigma}$ where $\gamma_\sigma = \alpha_\sigma -
\beta_\sigma$.  Using the Keldysh [17] formalism we obtain the
current:
\begin{equation}
I_\sigma = \frac{e}{2} (\frac{-i}{\hbar})^2 i \langle\langle
\int_0^t d t_1 \{ h_T(t_1 - i \varepsilon ) \frac{d}{d
\gamma_\sigma} h_T(t) - (\frac{d}{d \gamma_\sigma}h_T(t) ) h_T(t_1
+ i \varepsilon) \} \rangle\rangle
\end{equation}
where $ \langle\langle \; \rangle\rangle$ stands for the
thermodynamic average at temperature $T$ with respect to the
leads: $H_L + H_R + \mu_L( N_\uparrow + N_\downarrow) + \mu_R(
n_\uparrow + n_\downarrow)$.  We introduce the drain source
voltage $V_{DS} = \frac{\mu_L - \mu_R}{e}$ and the gate voltage
$V_G = \frac{\mu_L + \mu_R}{2 e}$.  The chemical potentials
$\mu_L$ and $\mu_R$ are used to perform the thermodynamic average
with respect the ''zero modes'' in the leads.  The current is
controlled by the short wire correlation function
$K_\sigma(t-t_1)$ (see eq. 6e).

Using eq. 7 we can compute the tunneling conductance.  Changing
the gate voltage $V_G$ we can reach a situation that $|E_\uparrow
- V_G| \ll \hat{\lambda}^2$ and $|E_\downarrow - V_G|
> \hat{\lambda}^2$.

As a result $E_\downarrow$ is off-resonance and $E_\uparrow$ is at
resonance. Since the conductance in eq. 7 depends on
$K_\sigma(t-t_1)$ given by eq. 6e we have two different situations
for the two conductances $G_\uparrow$ and $G_\downarrow$.  For
$\sigma = \downarrow$, the off-resonance condition for
$E_\downarrow$ allows to replace $K_\downarrow(t-t_1)$ by
$\delta(t-t_1)$.  This gives rise in eq. 7 to ``weak link''
problem in a Luttinger liquid.  As a result we find that
$G_\downarrow$ is given by $G_\downarrow \sim \frac{e^2}{h}
(\frac{T}{T_F})^{2 (r-1)}$, $r > 1$.  Therefore lowering the
temperature causes $G_\downarrow$ to decrease towards zero.

For spin up, we have a resonance condition. $E_\uparrow$ is at the
Fermi level, therefore $K_\uparrow (t-t_1)$ is replaced by
$K_\uparrow (t-t_1) \sim 1$.  Substituting $K_\uparrow (t-t_1)
\sim 1$ in eq. 7 introduces a long time correlation expressed
mathematically by a shift in the exponent $r \rightarrow r-1$.  As
a result, we find in this case that the conductance $G_\uparrow
\sim \frac{e^2}{h} (\frac{T}{T_F})^{2 (r - 2)}$.  Contrary to the
previous case here we observed that by lowering the temperature
$G_\uparrow$ increases towards the maximal value $G_\uparrow \sim
\frac{e^2}{h}$.  Since $G_\downarrow \sim 0$ we find that
$G_\uparrow + G_\downarrow \geq e^2/h$ in agreement with the
experimental situation.

(b) Next we consider the situation in the limit $T < T^{wire}$. In
this case we stop scaling at the length scale $b = d/a$.  At this
scale we have a ''zero dimension'' quantum problem.

At the length scale $b = \frac{d}{a}$, $\ell = \log(d/a)$ the
bosonic fields $\Phi_{c(s)}(x)$ are integrated out and eqs. 3a and
3b are replaced by the zero mode hamiltonians $H_c^{(n=0)}$ and
$H_s^{(n=0)}$.  (The zero mode representation is valid at low
temperature T such that the thermal length $L_T \equiv \frac{
\hbar v_F}{K_B T} > d \equiv \frac{ \hbar v_F}{K_B T^{wire}}$ (For
wires of the length $ d \sim \mu m$ we obtain temperatures, $T
\sim 1 - 2 K^\circ$) Therefore for temperatures $T < T^{wire}$) we
replace eqs 3a and 3b only by the zero mode hamiltonian $H_c^{(n =
0)} = \varepsilon_c h_c$ and $H_s^{(n = 0)} = \varepsilon_s h_s$
where $\varepsilon_c = \frac{\hbar \pi}{2 d} v_c$ and
$\varepsilon_s = \frac{\hbar \pi}{2 d} v_s$.

In the quantum region $P_c$ and $P_s$ are discrete and if a
non-zero value of $\hat{\mu}_s$ occurs it must be an integer.  But
in this case the partition function is invariant if we shift $P_s
\rightarrow P_s \pm 1$ we conclude that no broken symmetry takes
place, $Z(\hat{\mu}_s ) = Z(0)$. The charge part is given by:
\begin{subequations} \label{eq8:whole}
\begin{equation}
h_c = \frac{1}{2} P_c^2 + g_c \cos(\pi P_c) \label{subeq8:1}
\end{equation}
$g_c = g_c(\delta, d)$, at half filling $\delta = 0$, $g_c \simeq
U/ \pi v_c$; $g_c \neq 0$ only for $\delta d < \pi$.\\

The spin part is given by:
\begin{equation}
h_s = \frac{1}{2} (P_s)^2 + g_s \cos(\pi P_s) \label{subeq8:2}
\end{equation}
\end{subequations}
$g_s \simeq \frac{\hat{U}}{1 + \hat{U} \ell}$, $\hat{U} = U/ \pi
v_s$, $\ell = \ln(d/a)$.  Therefore away from half filling we will
have $g_s \gg g_c$. \\

We want to compute the spectrum of the hamiltonian in eqs 8a-8b.
The spectrum of the free part of the hamiltonian in eqs 8a-8b is
given by $ | :\delta N_\uparrow :, : \delta N_\downarrow : \rangle
= |P_c\rangle \otimes | P_s \rangle $ where $P_c$ is the charge
sector and $P_s / 2 $ is the spin sector ($ : \delta N_{\uparrow,
\downarrow} : \equiv N_{\uparrow, \downarrow} - \langle
N_{\uparrow, \downarrow} \rangle$)  The low energy particles
excitation (which do not include the bosonic particle hole
excitation) obey the condition: $P_c = P_s$ (modulo 2), where $P_s
= 0$ corresponds to a singlet and $P_s/2 = \pm 1/2$ to the spin
half doublet.  Therefore the non-interacting spectrum will be
given by $|P_c = 2n \rangle \otimes | P_s = 0 \rangle$ and $|P_c =
2n+1 \rangle \otimes | P_s = \pm 1 \rangle$ where $n = 0, \pm 1,
\pm 2, \ldots$. The spectrum in the quantum case for the charg
part is given by $E_c(2n) \approx \frac{1}{2} \epsilon_c (2 n)^2$
and $E_c(2n+1) \approx \frac{1}{2}\epsilon_c (2n+1)^2$ and for the
spin part we have $E_s(0) = \epsilon_s g_s$ and
$E_s(\pm 1) = (\frac{1}{2} - g_s) \epsilon_s$. \\

At length scale $\ell = \ln{(d/a)}$ the creation and annihilation
fermion operators for the wire are replaced by the zero mode part,
$\chi_\sigma(x) \rightarrow \hat{\chi}_\sigma \equiv \frac{e^{i
q_\sigma}}{\sqrt{2 \pi a}}$.  In order to compute the tunneling
current we need the short wire correlation functions, $\langle
\hat{\chi}_\sigma(t) \hat{\chi}_\sigma^\dagger (t_1) \rangle
\equiv \frac{1}{2 \pi a} K_\sigma(t-t_1)$.  This can be done with
the help of the spectrum of the short wire.
\begin{eqnarray}
e^{i q_\uparrow} &=& \sum_{P_c = \{even\}} \left[ |P_c \rangle
\langle P_c + 1 | \otimes | P_s = 0 \rangle \langle P_s = 1 | + |
P_c -1 \rangle \langle P_c | \otimes | P_s = -1 \rangle \langle
P_s = 0 | \right]  \\
e^{i q_\downarrow} &=& \sum_{P_c = \{even\}} \left[ |P_c \rangle
\langle P_c + 1 | \otimes | P_s = 0 \rangle \langle P_s = -1 | + |
P_c -1 \rangle \langle P_c | \otimes | P_s = 1 \rangle \langle P_s
= 0 | \right]
\end{eqnarray}
Using the spectrum of $h_c$ and $h_s$ we compute $K_\sigma(t-t_1)$
using eqs 9, 10.  At low temperature case (b) $T < T^{wire}$ we
find:
\begin{equation}
K_\sigma(t-t_1) = (1 + 2 e^{- \beta \Delta})^{-1} \left[ e^{i
\Delta (t-t_1)} + e^{- \beta \Delta} e^{- i \Delta (t -t_1)}
\right]
\end{equation}
where $\Delta$ is given by
\begin{equation}
\Delta = E_c(\pm1) - E_c(0) + E_s(\pm 1) - E_s(0) = \frac{1}{2} (
\varepsilon_c + \varepsilon_s) -  \frac{1}{2} g_s \varepsilon_s
\end{equation}
In the limit of $\beta \rightarrow \infty$ we have a resonant
impurity problem of energy $\Delta$ above the Fermi energy.\\

Comparing eq. 11 with eq 6f we observe that eq 11 is spin
independent.  Therefore by tuning the gate voltage we obtain a
resonant impurity problem wich obeys $G_\uparrow \sim G_\downarrow
\sim \frac{e^2}{h} (T/T_F)^{2 ( r - 2)}$.  At low temperatures
$G_\sigma$ grows reaching the maximum value $G_\uparrow +
G_\downarrow \simeq \frac{2 e^2}{h}$. \\

In conclusion we have shown that a system consisting of two
Luttinger leads coupled through a short wire with e-e interaction
has the following behavior: at temperature $T > T^{wire}=
\frac{\hbar v_F}{k_B d}$ the short wire acts as a spin polarizer
with a conductance $G \geq \frac{e^2}{h}$.  From the other hand at
$T < T^{wire} $ the system is equivalent to a resonant impurity
level with a conductance $G \simeq \frac{2 e^2}{h} $ at $T
\rightarrow 0$.  Therefore we suggest that our system can work as
a spin filter by varying the temperature.  This avoids the
difficult practical task of applying a magnetic field on the short
wire.

\begin{center} ACKNOWLEDGEMENT \end{center}

This work has been supported by the DOE Grant No.
DE-FG02-01ER4909.


\end{document}